\documentclass{astlb}

\usepackage{graphicx}
\usepackage{color}
\usepackage[hyperfootnotes=false]{hyperref}

\definecolor{darkblue}{rgb}{0,0,0.9}

\def\smfigure#1#2#3{
  \begin{minipage}{1.0\columnwidth}
    \begin{minipage}{0.049\columnwidth}
      \rotatebox{90}{\footnotesize\phantom{0000}#3}
    \end{minipage}
    \begin{minipage}{0.9\columnwidth}
     \includegraphics[viewport=40 188 556 678,width=0.97\columnwidth]{#1}
      \centerline{\footnotesize #2}
    \end{minipage}

    \vskip 3pt
    ~
  \end{minipage}
}

\def\smfiguresmall#1#2#3{
  \begin{minipage}{0.95\columnwidth}
    \begin{minipage}{0.049\columnwidth}
      \rotatebox{90}{\footnotesize\phantom{0000}#3}
    \end{minipage}
    \begin{minipage}{0.95\columnwidth}
      \includegraphics[viewport=40 188 556 540,width=0.97\columnwidth]{#1}
      \centerline{\footnotesize #2}
    \end{minipage}

    \vskip 3pt
    ~
  \end{minipage}
}

\def\ps1{\emph{Pan-STARRS1}}

\def\cl2305{SRG\lowercase{e}\,CL2305.2$-$2248}

\usepackage{amssymb,stackengine,graphicx,scalerel}

\newcommand\vargtrsim{\mathrel{\ensurestackMath{\ThisStyle{%
  \stackengine{-.4\LMex}{\SavedStyle>}{%
    \rotatebox{25}{$\SavedStyle\sim$}}{U}{l}{F}{T}{S}}}}}

\newcommand{\ec}[1]{{\color{red} #1}}

\begin{document}

\journalinfo{2021}{0}{0}{1}[0]

\title{Observation of a very massive galaxy cluster at
  \lowercase{$z=0.76$} in SRG/eROSITA all-sky survey}

\author{R.A.~Burenin\email{rodion@hea.iki.rssi.ru}\address{1},
  I.F.~Bikmaev\address{2,3},
  M.R.~Gilfanov\address{1,4},
  A.A.~Grokhovskaya\address{5},
  S.N.~Dodonov\address{5},
  M.V.~Eselevich\address{6},
  I.A.~Zaznobin\address{1},
  E.N.~Irtuganov\address{2,3},
  N.S.~Lyskova\address{1},
  P.S.~Medvedev\address{1},
  A.V.~Meshcheryakov\address{1},
  A.V.~Moiseev\address{1,5},
  S.Yu.~Sazonov\address{1},
  A.A.~Starobinsky\address{7},
  R.A.~Sunyaev\address{1,4},
  R.I.~Uklein\address{5},
  I.I.~Khabibullin\address{1,4},
  I.M.~Khamitov\address{2,8},
  E.M.~Churazov\address{1,4}
  \addresstext{1}{Space Research Institute RAS (IKI), Moscow, Russia}
  \addresstext{2}{Kazan Federal University, Kazan, Russia}
  \addresstext{3}{Academy of Sciences of The Republic of Tatarstan, Kazan, Russia}
  \addresstext{4}{Max Planck Institute for Astrophysics, Garching, Germany}
  \addresstext{5}{Special Astrophysical Observatory of the Russian Academy of Sciences, Nizhnij Arkhyz, Russia}
  \addresstext{6}{Institute of Solar-Terrestrial Physics SB RAS, Irkutsk, Russia}
  \addresstext{7}{Landau Institute for Theoretical Physics RAS, Chernogolovka, Russia}
  \addresstext{8}{TÜBİTAK National Observatory, Antalya, Turkey}
}

\shortauthor{Burenin et al.}

\shorttitle{Distant and massive cluster from SRG/eROSITA survey}


\submitted{June 6, 2021}

\begin{abstract}
  The results of multiwavelength observations of the very massive
  galaxy cluster \cl2305 detected in X-rays during the first
  SRG/eROSITA all-sky survey are discussed. This galaxy cluster was
  also detected earlier in microwave band through the observations of
  Sunyaev-Zeldovich effect in South Pole Telescope
  (SPT-CL\,J2305$-$2248), and in Atacama Cosmological Telescope
  (ACT-CL\,J2305.1$-$2248) surveys. Spectroscopic redshift
  measurement, $z=0.7573$, was measured at the Russian 6-m BTA
  telescope of the SAO RAS, in good agreement with its photometric
  estimates, including a very accurate one obtained using machine
  learning methods. In addition, deep photometric measurements were
  made at the Russian-Turkish 1.5-m telescope (RTT150), which allows
  to study cluster galaxies red sequence and projected galaxies
  distribution. Joint analysis of the data from X-ray and microwave
  observations show that this cluster can be identified as a very
  massive and distant one using the measurements of its X-ray flux and
  integral comptonization parameter only. The mass of the cluster
  estimated according to the eROSITA data is
  $M_{500}=(9.0\pm2.6)\cdot10^{14}\, M_\odot$.  We show that this
  cluster is found among of only several dozen of the most massive
  clusters in the observable Universe and among of only a few the most
  massive clusters of galaxies at $z>0.6$.
  
  
\end{abstract}

\section{Introduction}

It is expected that in the all-sky X-ray survey which will be carried
out by eROSITA telescope on board the Spectrum Roentgen Gamma (SRG)
observatory, of order of 100\,000 of galaxy clusters will be detected,
including \emph{all} clusters with masses above
$M_{500}\approx 3\cdot 10^{14} M_\odot$ in the observable Universe
\citep{2021arXiv210413267S, 2021A&A...647A...1P}. This sample will be
of great interest for various cosmological studies. For example, these
data will allow to obtain new constraints on the parameters of the
cosmological model using the measurements of galaxy clusters mass
function
\citep[e.g.,][]{2009ApJ...692.1060V,2014A&A...571A..20P,2016A&A...594A..24P}.
With these data it will be possible to measure $\sigma_8 (z)$
dependence, which in turn will give an independent method to measure
the dependence of the Hubble parameter $H (z)$ on the redshift
\citep[e.g.,][]{2020MNRAS.494..819L}.

The SRG all-sky survey was started in December 2019, after the
successful launch of the SRG Space Observatory in July of the same
year. Currently, two complete all-sky surveys have been carried out
with the telescopes onboard SRG observatory, allowing to detect all
the most massive clusters of galaxies in the observable Universe, with
masses above $M_{500}\approx 6\cdot 10^{14} M_\odot$, located at
redshifts up to $z\approx 1$.

Galaxy clusters of that high masses are also observed in \emph{Planck}
all-sky survey \citep[\emph{PSZ2},][]{PSZ2}. However, it turns out
that not all very massive clusters at high redshifts are included in
\emph{PSZ2} catalog. Some of these clusters were later identified in
the optical among the SZ sources from \emph{PSZ2} catalog with no
optical identification \citep[e.g.,][]{br18_pazh, Z19_pazh}. In
addition, as we show below, SZ sources associated with very massive
galaxy clusters may not be detected in \emph{Planck} survey above
\emph{PSZ2} catalog threshold.

In this paper we discuss the observation of a very massive galaxy
cluster \cl2305, which was detected in SRG/eROSITA all-sky
survey. Using optical observations at the Russian-Turkish 1.5-m
Telescope (RTT-150) and at the 6-m telescope of SAO RAS (BTA), deep
photometric data for the field of this cluster, as well as
spectroscopic redshift measurement were obtained. Previously, this
cluster was also observed in the South Pole Telescope SZ survey
\citep{2020ApJS..247...25B} and Atacama Cosmology Telescope
\citep{2021ApJS..253....3H} Sunyaev-Zeldovich surveys. The galaxy
cluster \cl2305\ mass estimates obtained by different methods are in
good agreement with each other. It is shown that this cluster is one
of only a few most massive clusters in the observable Universe at a
redshifts $z>0.6$.

\begin{figure}
  \centering

  \includegraphics[viewport=10 147 565 706,width=\columnwidth]{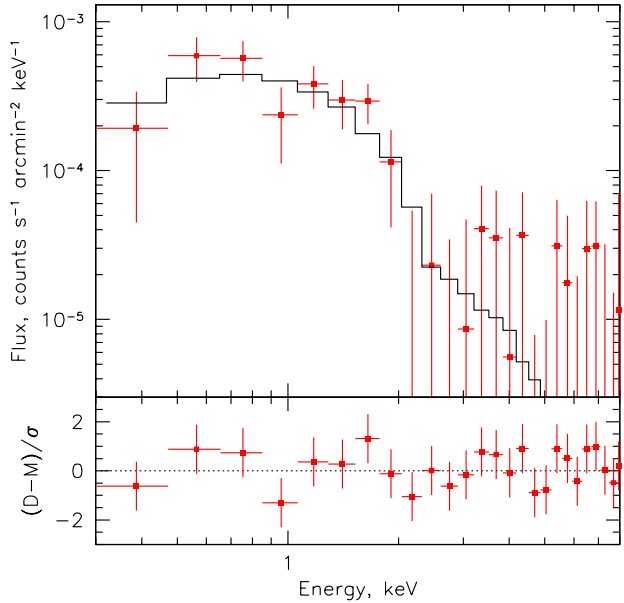}
  
  \vskip -2mm
  \caption{The spectrum of an extended X-ray source \cl2305\ obtained
    in SRG/eROSITA survey. The solid curve shows the model spectrum
    --- radiation of an optically thin plasma with temperature
    $kT=11$~keV. The statistical significance of the data allows to
    exclude the temperature below 3~keV, at the level of
    2~$\sigma$. The flux is normalized for one of the 7 modules
    of the eROSITA telescope.}
  \label{fig:ero_spec}
  
\end{figure}

\begin{figure*}
  \centering

  \includegraphics[width=0.9\columnwidth]{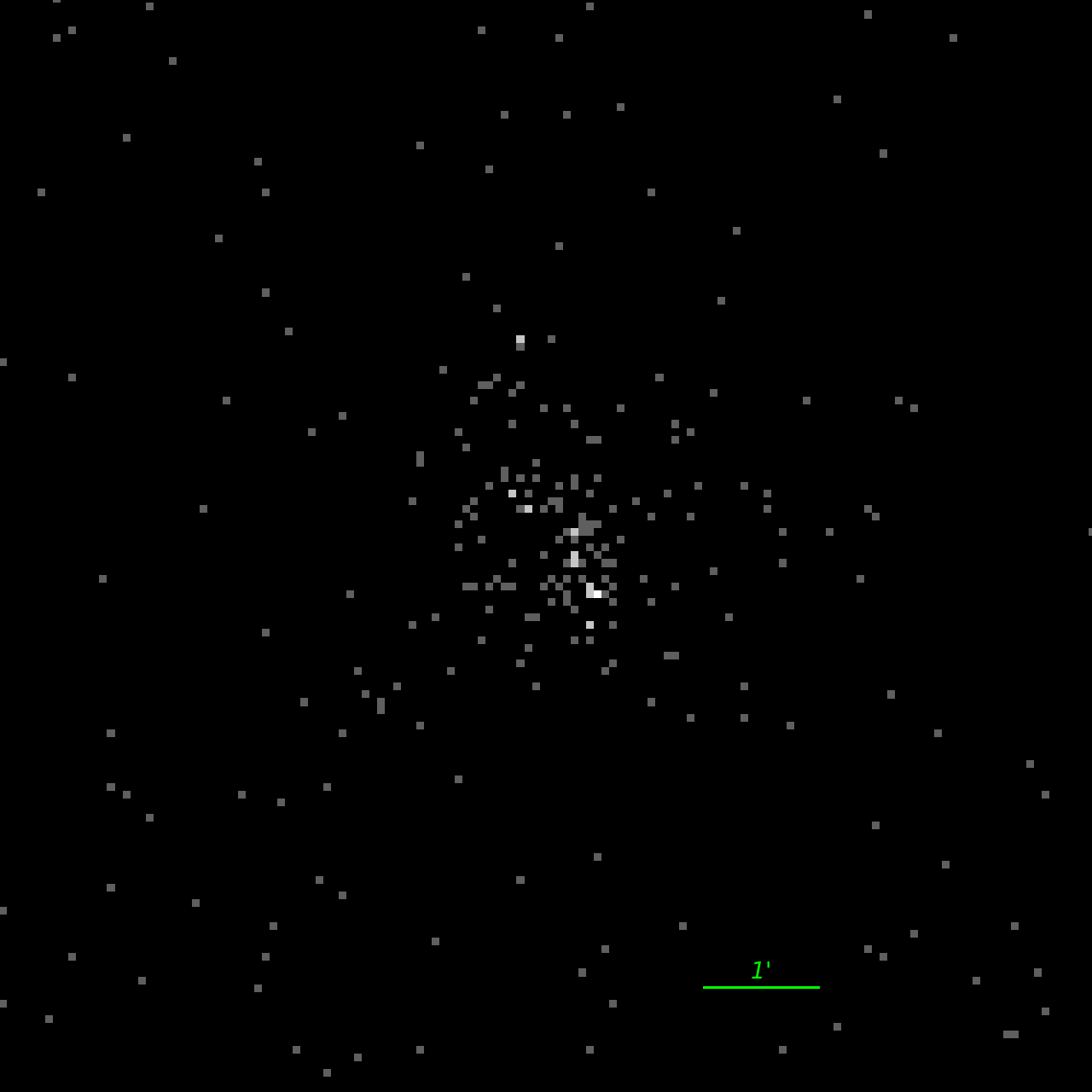}
  \includegraphics[width=0.9\columnwidth]{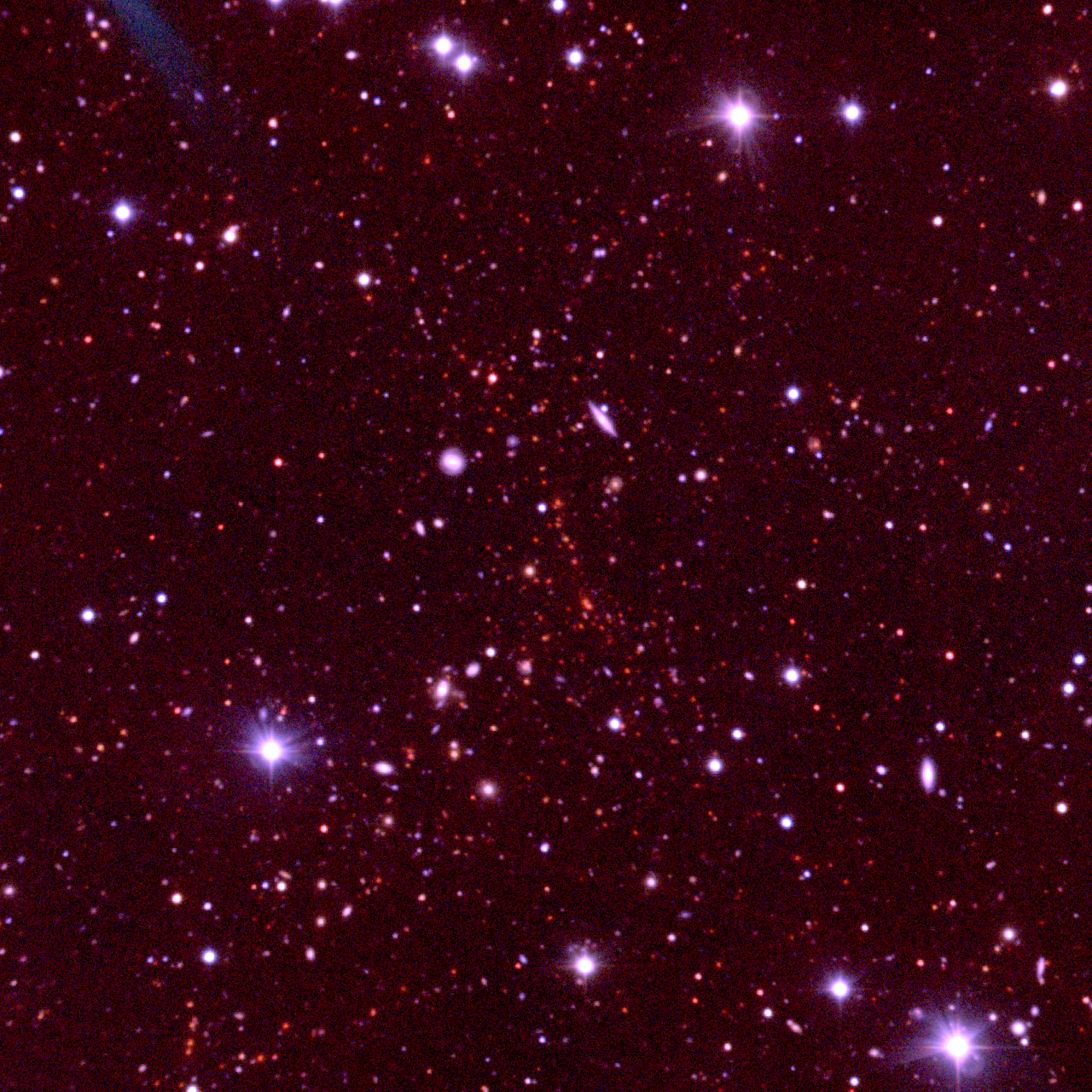}

  \includegraphics[width=0.9\columnwidth]{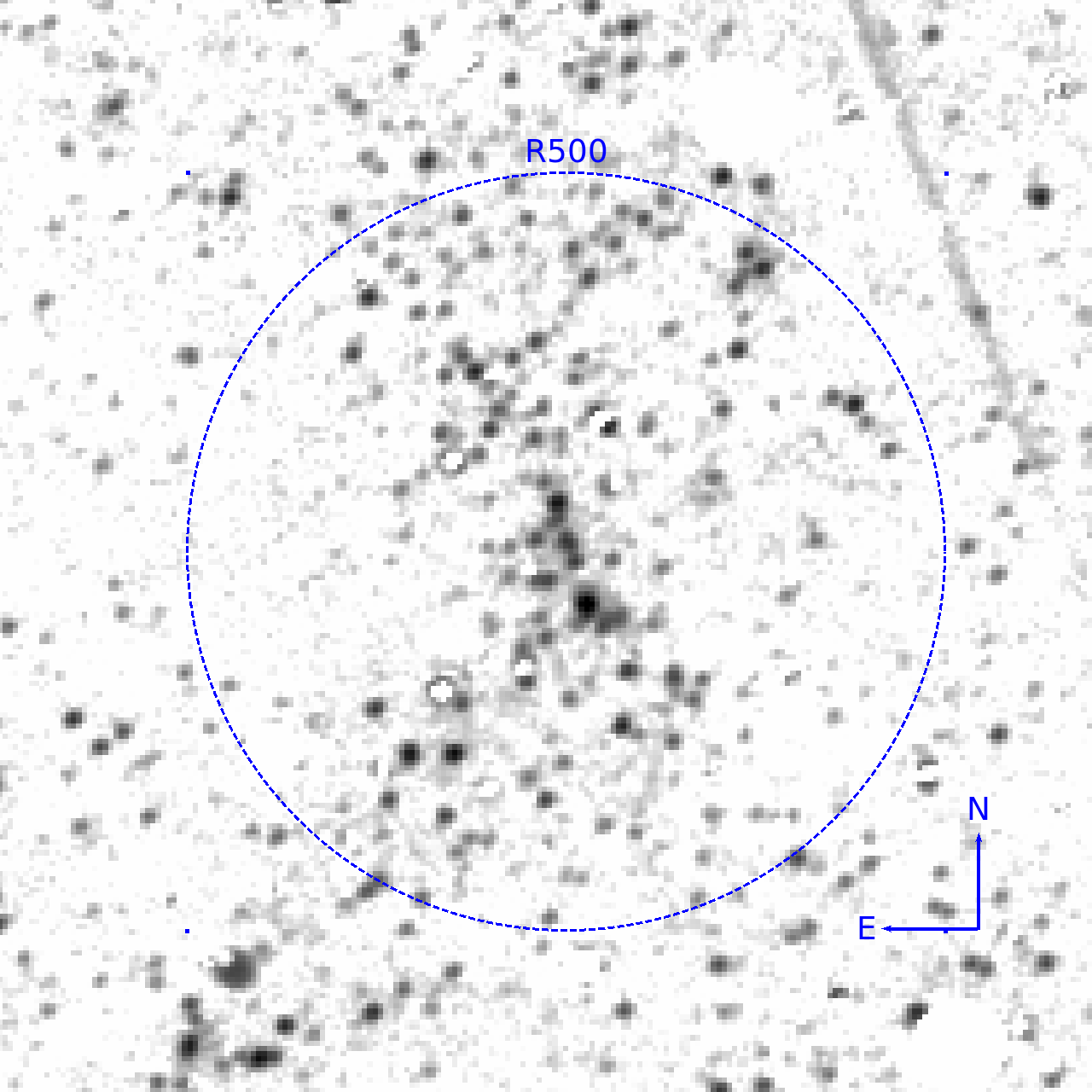}
  \includegraphics[width=0.9\columnwidth]{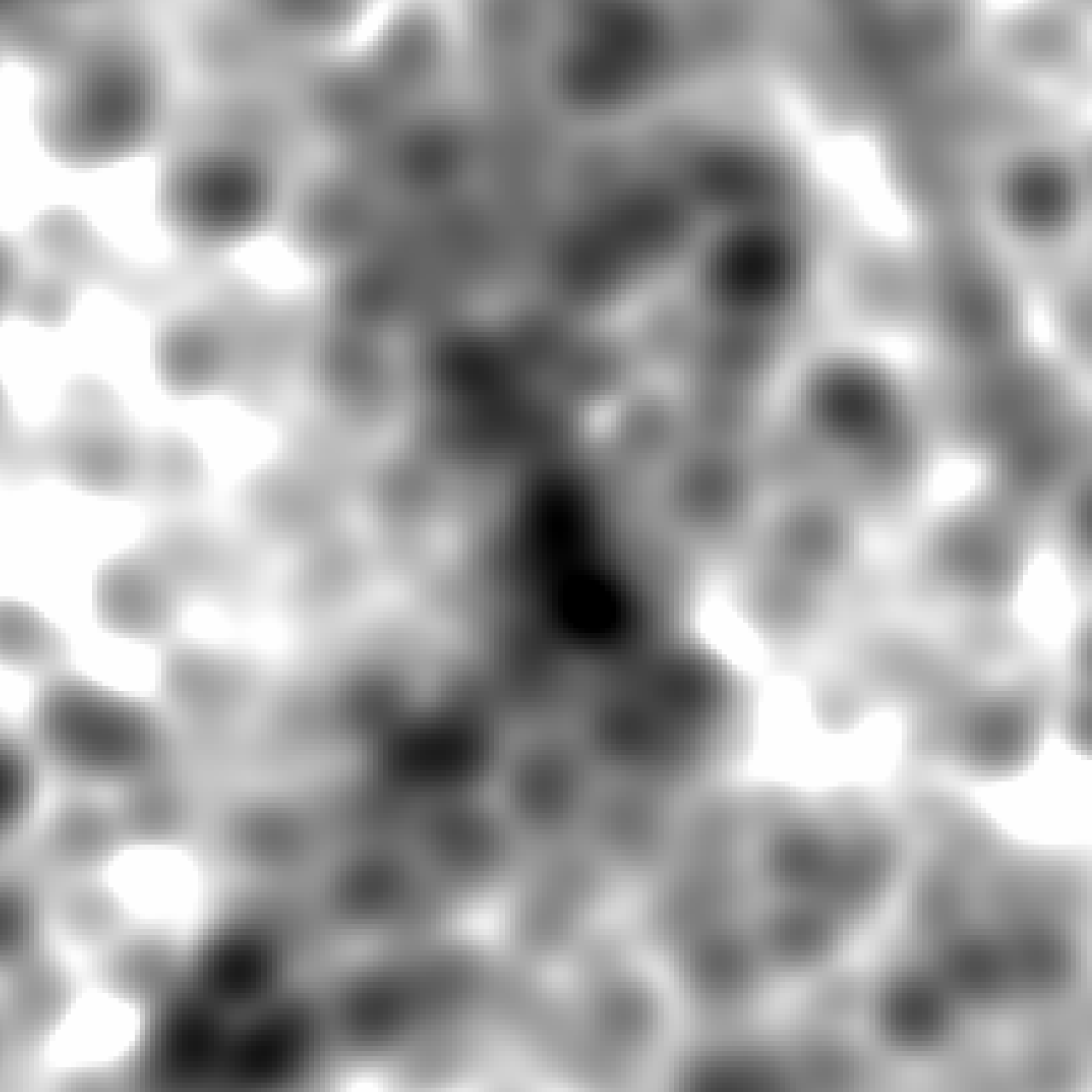}

  \bigskip

  \caption{X-ray image of the field of the galaxy cluster \cl2305\
    obtained in SRG/eROSITA survey (top, left), pseudo-color image in the filters \emph{i, r, g} \emph {(RGB)} obtained on RTT-150 telescope with about $1\arcsec\!.5$ seeing (top, right), the \emph{WISE} image in the 3.4~$\mu$m band, cleaned from stars and galaxies that are not belong to the cluster red sequence (bottom,
    left), the same image, smoothed by a two-dimensional Gaussian,
    $\sigma=8\arcsec$ (bottom, right). The image at the bottom left
    shows a circle with a radius of $R_{500}=3\arcmin\!.3$. All images
    are presented at the same scale, the field size is
    $9\arcmin\!\!.4\times9\arcmin\!\!.4$, the center of the field
    coincides with the X-ray center of the cluster.}
  
  \label{fig:images}
  
\end{figure*}

\begin{figure}
  \centering
  
  \smfigure{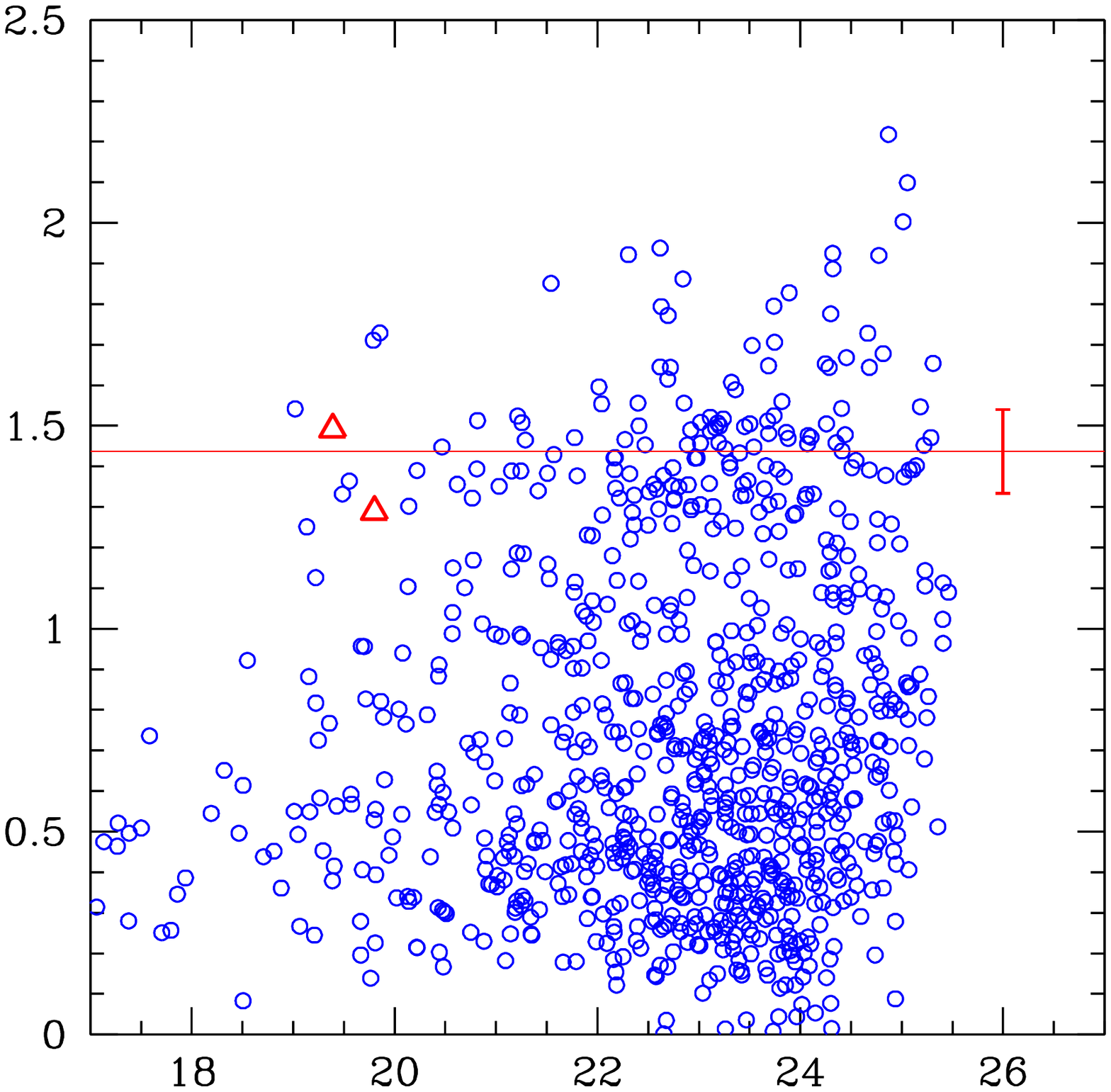}{\emph{i}}{\emph{r$-$i}}

  \caption{Color-magnitude diagram for galaxies located within
    $3\arcmin\!.3$ radius from the X-ray center of the cluster
    \cl2305. The position of the red sequence of cluster galaxies is
    marked by a horizontal line, the vertical error bar shows the
    characteristic size of the color measurement error for each individual galaxy at $\approx 23^m$. The red triangles indicate the central cluster galaxies, for which spectroscopic redshift measurements were obtained.}
  \label{fig:rs}

\end{figure}

\section{Observations in SRG/\lowercase{e}ROSITA survey}

As of the beginning of June 2021, the field of the galaxy cluster
\cl2305\ was observed in SRG/eROSITA all-sky survey for three times
--- in June and December 2020, as well as in May 2021. The full
exposure, corrected for vignetting, is currently about 250~s. The data
of the eROSITA telescope was processed using the
\emph{eSASS}\footnote{https://erosita.mpe.mpg.de/edr/DataAnalysis/}
software using pre-flight calibration data. The X-ray image of the
cluster field \cl2305\ in 0.5--2~keV energy range is shown in
Fig.~\ref{fig:images} in the top left panel. In total, about 140
photons were observed from this source during the first year of the
SRG/eROSITA survey.

The sources detection in the all-sky survey was carried out using the
wavelet decomposition of X-ray images \citep{1998ApJ...502..558V} and
the calculation of various characteristics of sources was carried out
by the maximum likelihood fitting using \emph{ermldet} task from the
\emph{eSASS} package. In this field, in the all-sky survey, previously
unknown extended X-ray source at the coordinates
$\alpha=23^h05^m11^s.6, \delta=-22^d48^m54^s$ was discovered. Under
the assumption that the radial profile of the surface brightness of an
extended source is described by the $\beta$-model:
$I=I_0 (1+r^2/r_c^2)^{-3\beta+0.5}$ \citep{1976A&A....49..137C}, with
$\beta=2/3$, the significance of the source extention corresponds to
$\delta\chi^2=30.66$, i.e.\ is high, the radius of the $\beta$-model
is $r_c=20\arcsec\!\!.1$.  The source's X-ray flux in 0.5--2~keV
energy range, calculated using \emph{ermldet}, corrected for
interstellar Galactic absorption with a column density
$N_H = 2.2 \cdot 10^{20}$~cm$^{-2}$, is
$6.62\pm 0.61 \cdot 10^{-13}$~erg~s$^{-1}$~cm$^{-2}$. Note that the
central part of the extended X-ray source is of irregular structure,
i.e.\ the hot gas distribution is probably disturbed, which may be the
result of the outflow of hot gas from the active nucleus of one of the
central galaxies in the cluster. On the other hand, this may also be a
consequence of large-scale gas motions.

In Fig.~\ref{fig:ero_spec} the spectrum of an extended X-ray source
obtained from the SRG/eROSITA survey data is shown. The spectrum was
extracted from a circle with a radius of $5\arcmin$ centered on the
cluster. The charged particles induced background was removed using
calibration observations with the detector closed by X-ray opaque
filter. The sky X-ray background in the direction of the cluster was
estimated in $5\arcmin$--$50\arcmin$ annulus around the cluster and
subtracted from the spectrum. The statistical significance of the data
does not allow to put an upper limit on the gas temperature, but
excludes a temperature below 3~keV, at the level of 2~$\sigma$. The
figure shows a an optically thin plasma spectrum model with a
temperature of 11~keV, which roughly corresponds to the
luminosity-temperature relation for a given redshift (see\ below).

\section{Observations in optical and IR}

\subsection{Direct imaging and photometry}

A massive cluster of galaxies was discovered in the field of an
extended X-ray source using our automated galaxy cluster red sequence
detection procedure based on optical and IR data from the
\emph{Pan-STARRS} \citep{2016arXiv161205560C} and \emph{WISE}
\citep{2010AJ....140.1868W} surveys. For that we use \emph{WISE}
survey coadds obtained from data for the full operation time of the
satellite \emph{WISE} \citep{2017AJ....154..161M}, versions of
\emph{NeoWISE-R6}, available online\footnote{http://unwise.me/}. We
used forced photometry based on these\emph{WISE} data for all objects
from the \emph{Pan-STARSS} survey, which was obtained using a complete
PSF model, which correctly takes into account its assymetry and wings
on large angular scales \citep{wise_ps1_forced}. The red sequence
detection was carried out using an automatic procedure similar to that
used by \cite{br17_pazh}, which gives the photometric redshift
estimate based on the red sequence colors,
$z_{\mbox{\scriptsize phot}} = 0.70\pm0.04$.

Deep direct images of the field of galaxy cluster \cl2305\ were
obtained on the Russian-Turkish 1.5-m Telescope (RTT-150) in the
period from August 23 to 26, 2020 using SDSS \emph{g,r,i,z}
filters. The total exposures were 13200~s, 13800~s, 8400~s and 8400~s,
for \emph{g,r,i,z} respectively, the image quality was about
$1\arcsec\!.5$. The total exposure in each filter was divided into
exposures of 600~s, between which the telescope pointing axis was
shifted by 10--20\arcsec\ in an arbitrary direction. The data
processing was carried out using the \emph{IRAF} package, as well as
using our own software. The standard set of calibrations was
applied. The pseudo-color image of this field in the filters
\emph{i,r,g} \emph {(RGB)} shown in fig.~\ref{fig:images} top
right. The red galaxies in the center of the image are cluster member
ones.

Photometric calibration of the images was obtained using observations
of photometric standards \citep{Smith02}, as well as by comparison
with photometric measurements from \emph{Pan-STARSS} survey. In
Fig.~\ref{fig:rs} the color--magnitude diagram is shown for galaxies
in the cluster field located within $3\arcmin\!.3$ radius from the
X-ray center of the cluster. To produce this diagram, objects catalog
obtained with the \emph{SExtractror} software
\citep{1996A&AS..117..393B} was used, starlike objects were excluded
($\mathrm{\mathtt{CLASS\_STAR}}>0.8$).  In the diagram we use
\texttt{MAG\_AUTO} magnitudes \citep{1980ApJS...43..305K}, the colors
were measured using aperture magnitudes with the apertures of the same
size, equal to the PSF FWHM. The cluster galaxies red sequence is
shown in Fig.~\ref{fig:rs}. It appears in Fig.~\ref{fig:rs} as a
\textit{concentraion} of dots near the red horizontal line. The dispersion of
colors near this line roughly corresponds to the photometric errors of
measuring the colors of galaxies. Due to the large uncertainties in
photometric measurements, it is impossible to obtain a reliable
measurement of the slope of the red sequence, as well as the internal
color dispersion.

In the lower panel of Fig.~\ref{fig:images} the images from
\emph{WISE} survey in $3.4\mu m$ band are shown. The models of all
stars and galaxies that are not member of cluster red sequence are
subtracted from these images. The image at the bottom right is
additionally smoothed by a two-dimensional Gaussian,
$\sigma=8\arcsec$. These images show the distribution of the cluster
red sequence galaxies in projection to the sky.

Fig.~\ref{fig:images} shows that in the central region of the cluster
of about arcminute size, the distribution of galaxies is elongated in
the direction of \emph{NNE--SSW}. At the redshift of the cluster (see\
below), this size corresponds to the size of about 500\,pc. At greater
distances from the center, the distribution of cluster galaxies in the
picture plane also remains inhomogeneous. In the East and West
directions, cluster galaxies are not observed at a distance larger
than an approximately arcminute from the center of the cluster, while
in the north and south directions, the surface density of cluster
galaxies remains significant up to distances of the order of and
greater than $R_{500}\approx 3\arcmin$ (see\ below). The shape of the
distribution of galaxies in the picture plane resembles an
hourglass. Such a strong central asymmetry of the distribution of
galaxies in the projection to the sky may indicate that the cluster is
not relaxed dynamically.

\subsection{Photometric redshift estimate using machine learning}

\begin{figure}
  \centering

  \smfigure{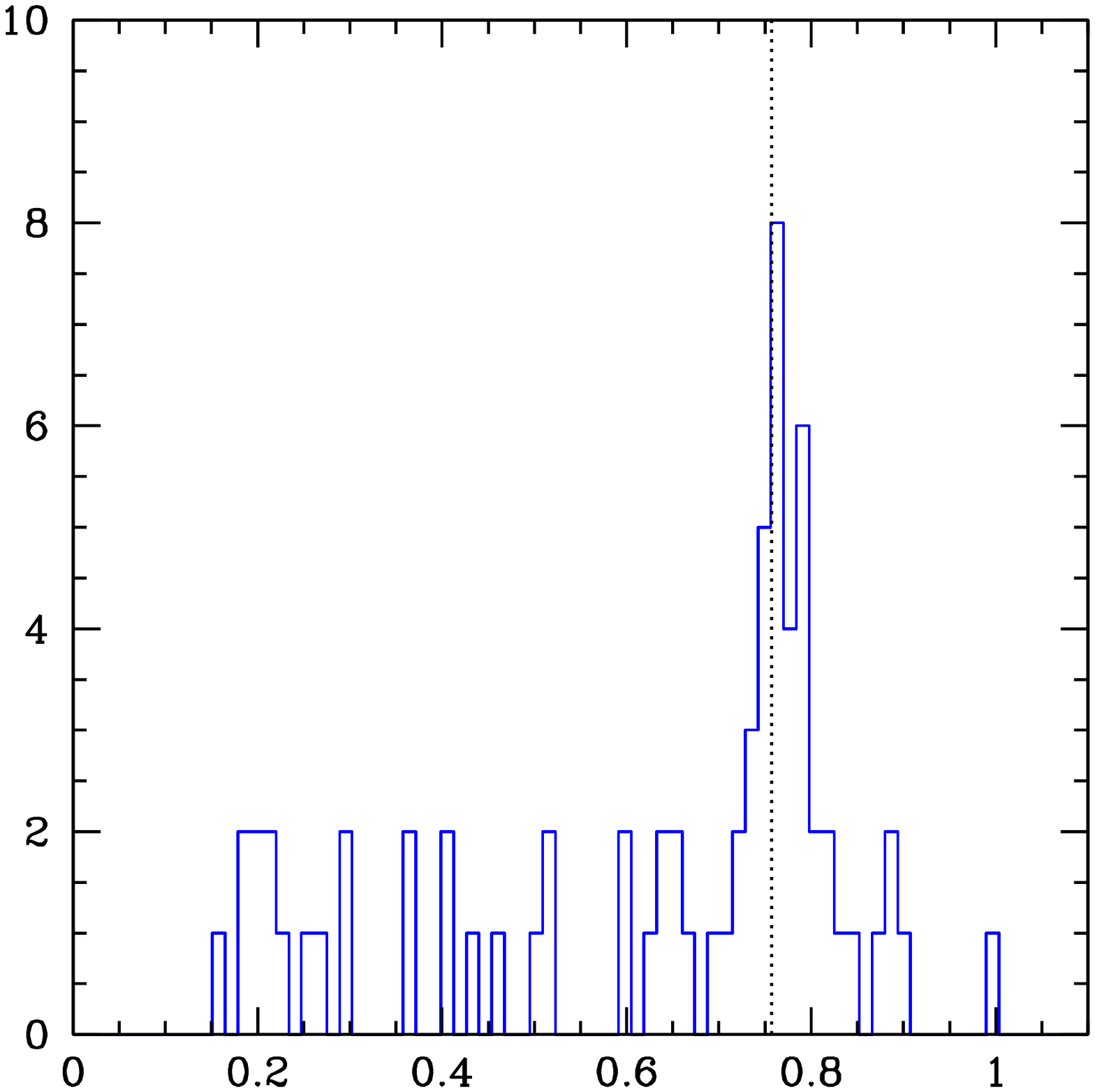}{\emph{z}}{\emph{N}}

  \caption{Distribution of galaxies photometric redshifts in the field
    of cluster \cl2305, located within $2\arcmin$ radius from the
    X-ray center of the cluster. The photometric redshift estimates
    are obtained from the \emph{Pan-STARRS} and \emph{WISE} surveys
    using machine learning methods. The spectroscopic redshift
    measurement, $z=0.7573$ is shown with vertical dotted line.}
  \label{fig:srgz_ds_z}
\end{figure}

Machine learning methods allow to measure photometric redshifts
(\emph{photo-z}) of galaxy clusters from the accurate photometric
redshift estimates for individual cluster galaxies. For example, the
accuracy of $\sigma\sim1$\% was obtained for clusters at redshifts
$z<0.45$ using SDSS data \citep{2015AstL...41..307M}. We have trained
a galaxies \emph{photo-z} model based on a quantile random forest
described in \cite{2018AstL...44..735M}. The photometric features for
the random forest model were the galaxies PSF and
\citep{1980ApJS...43..305K} magnitudes from \emph{Pan-STARRS1 DR2}
catalog, the \emph{W1} and \emph{W2} magnitudes obtained from
\emph{WISE} forced photometry in the corresponding filters
\citep{wise_ps1_forced}, as well as all possible colors based on these
optical and IR magnitudes. Using this photometric redshift estimates
model, point predictions were made for the redshifts of all galaxies
in the field of the cluster \cl2305{}. After that only the objects
with reliable \emph{photo-z} predictions were selected by the
confidence parameter $\texttt{zConf}>0.6$, where \texttt{zConf} is
defined in a standard way as the probability of the predictions of the
photometric estimate $z$ in the vicinity of
$z_{\mbox{\scriptsize phot}}\pm0. 06 (1+z_{\mbox{\scriptsize phot}})$.

In Fig.~\ref{fig:srgz_ds_z} we show the distribution of photometric
redhift estimates for the galaxies located within the $2\arcmin$
radius from the X-ray center of the cluster. The distribution has a
well-distinguishable peak, from which a photometric estimate of the
redshift of the cluster can be obtained:
$z_{\mbox{\scriptsize phot}} = 0.766\pm0.012$.

\begin{figure*}
  \centering
  
  \smfiguresmall{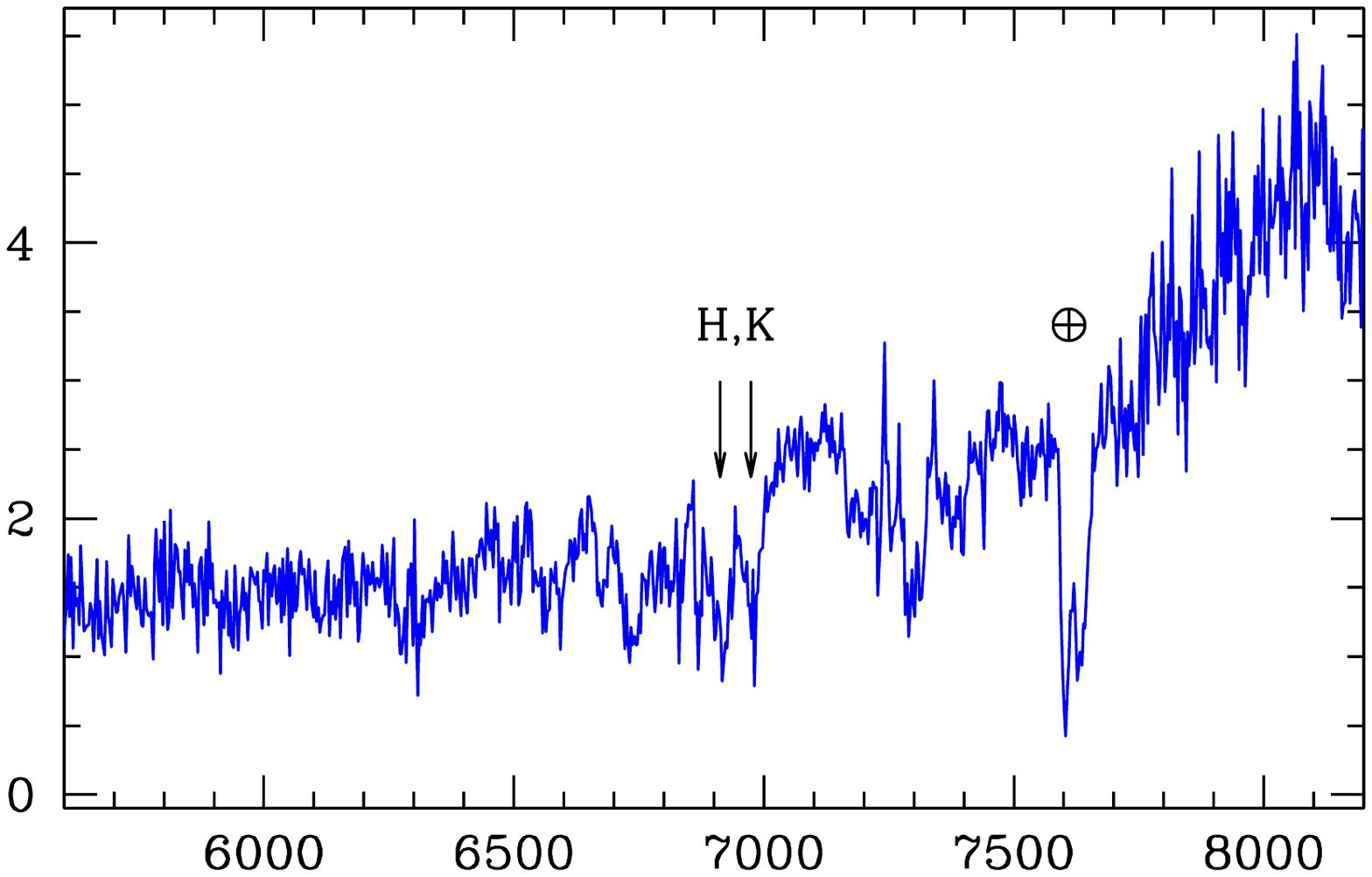}{wavelength, \AA}{flux, $10^{-17}$~erg\,s$^{-1}$\,cm$^{-2}$}
  \smfiguresmall{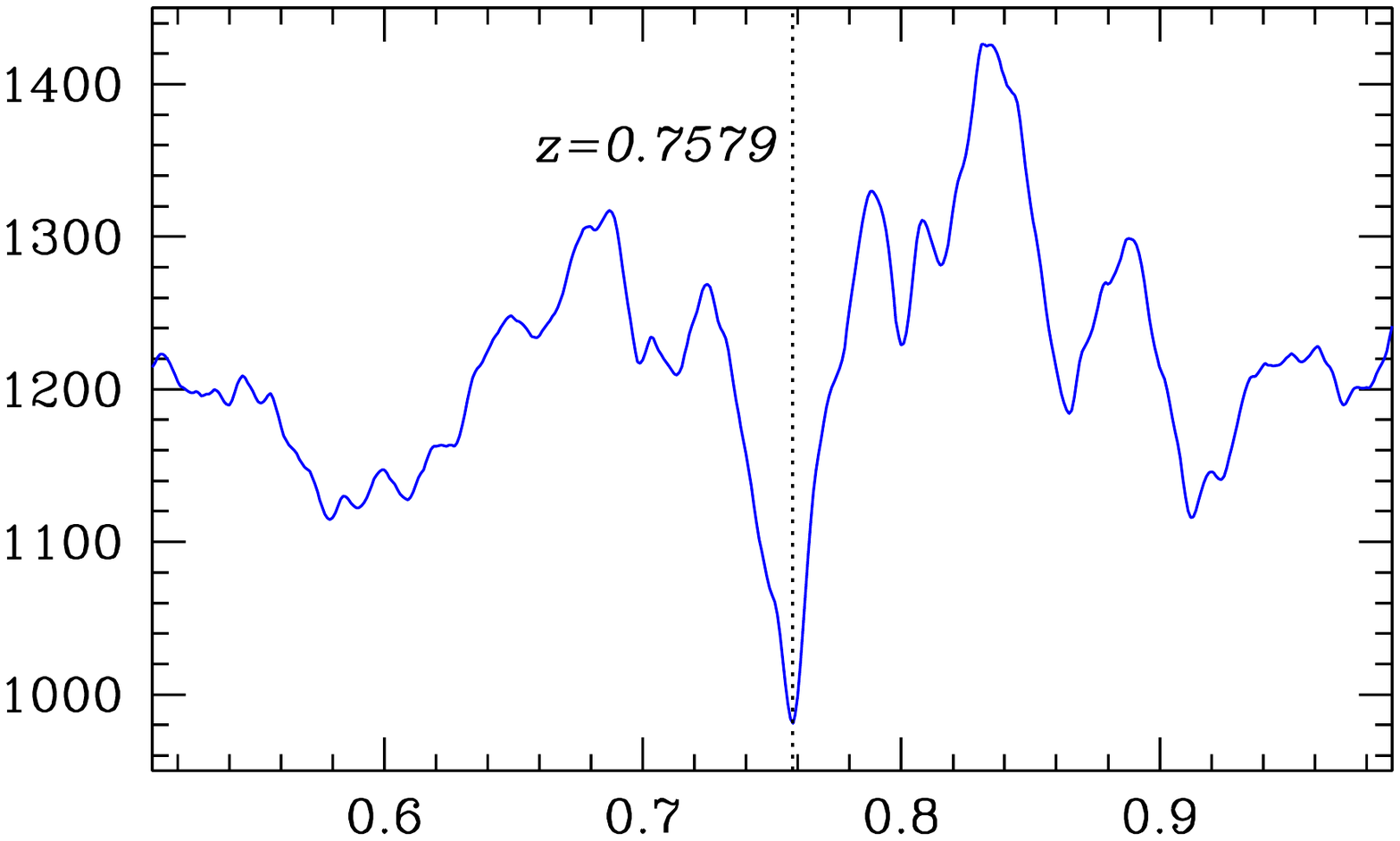}{$z$}{$\chi^2$}
  
  \smfiguresmall{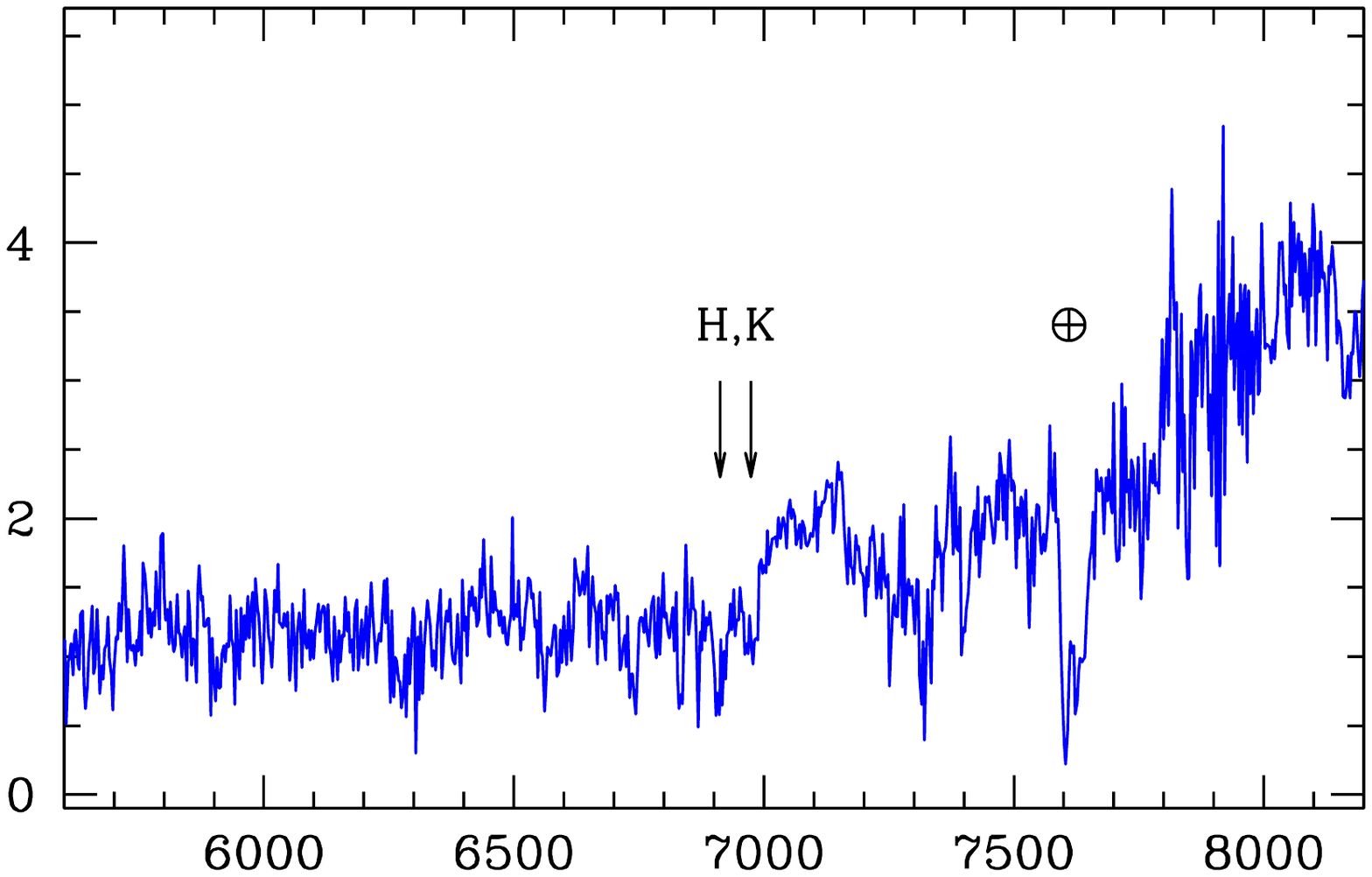}{wavelength, \AA}{flux, $10^{-17}$~erg\,s$^{-1}$\,cm$^{-2}$}
  \smfiguresmall{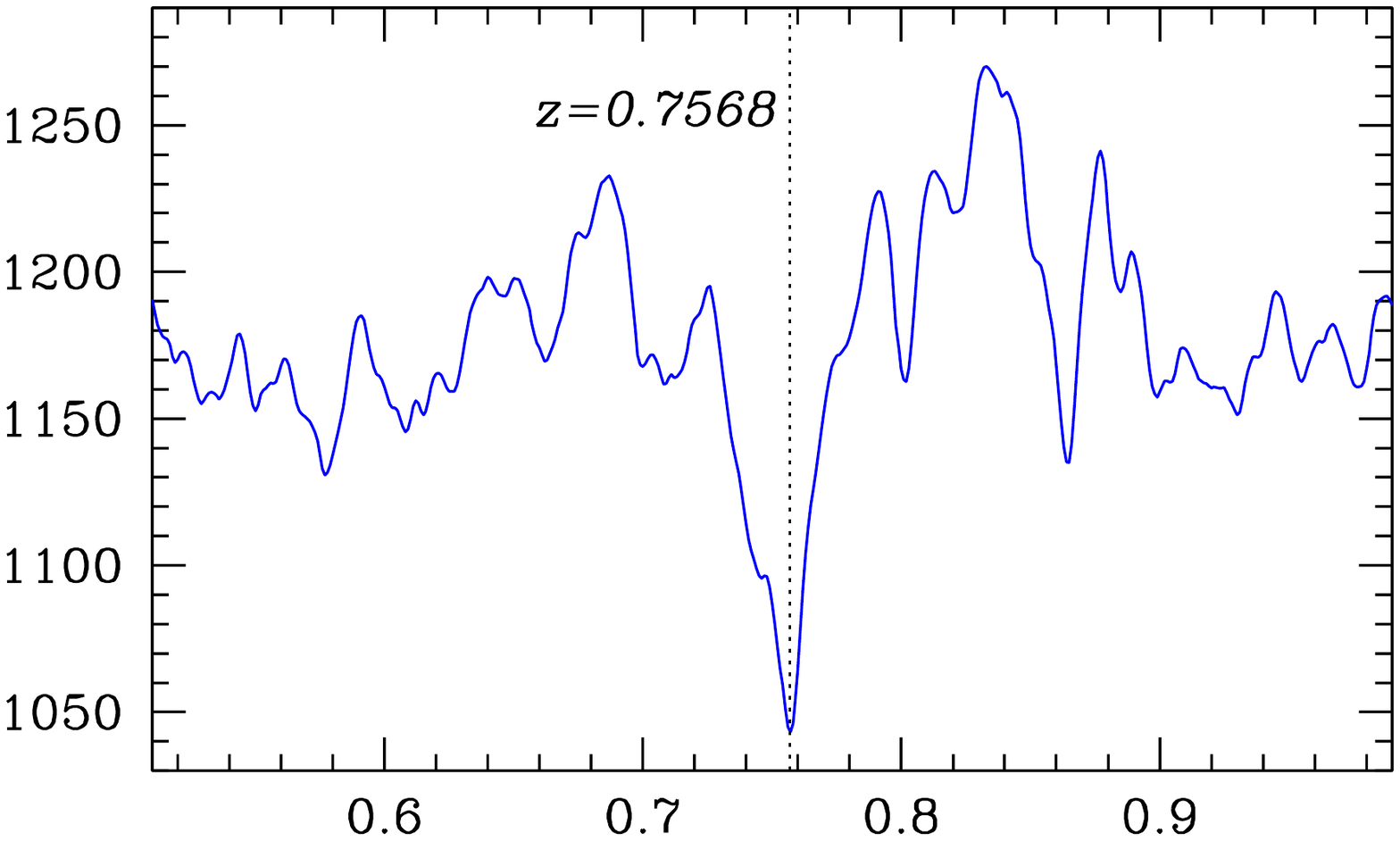}{$z$}{$\chi^2$}
  
  
  \caption{The spectra of the two central galaxies of the cluster
    \cl2305, obtained at the 6-m BTA telescope using the SCORPIO-2
    spectrograph (left), $\chi^2$ from the cross-correlation of the
    spectra with an elliptical galaxy template spectrum (right).}
  \label{fig:spec}
\end{figure*}

\subsection{Spectroscopy}

The spectra of the brightest cluster galaxies were obtained at the 6-m
BTA (Bolshoi Teleskop Alt-azimutalny) telescope of the Special
Astrophysical Observatory of RAS using the \emph{SCORPIO-2}
spectrograph \citep{scorpio11}. The observations were carried out
during the night of August 17, 2020. The positional angle of the long
slit was chosen so that two central galaxies with coordinates near
$\alpha=23^h05^m10^s.6, \delta=-22^d49^m11^s$, with magnitudes
$i=19.39$ and $i=19.80$, are observed in the slit. The slit width was
$1\arcsec\!.97$, an the seeing was $1\arcsec\!.8$. The spectrum was
obtained using the \emph{VPHG940@600} grism, in the spectral range
3500--8500\AA, with the spectral resolution of about 14\AA. The total
exposure time was 4800~s, it was divided into four exposures of
1200~s, between the exposures the objects were shifted for 14\arcsec\
along the slit.

The data were processed in a standard way, using the
\emph{IRAF} software package, as well as own software. The images of
two-dimensional spectra were bias and a flat field corrected, a
wavelength scale was set based on the spectra of calibration
lamps. Then two-dimensional spectra were aligned and combined,
one-dimensional spectra of galaxies and background spectra were
extracted from the two-dimensional spectrum in a standard way. The
calibration of the flux density in the spectra was done using
spectrophotometric observations of standards from the list of the
European Southern
Observatory\footnote{https://www.eso.org/sci/observing/tools/standards/}.

The spectra of the two brightest galaxies of the cluster are shown in
Fig.~\ref{fig:spec} on the left panels. Here, the absorption lines of
the calcium doublet \emph{H,K}, as well as a break near the wavelength
4000\AA\ are clearly visible. In the right panels of
Fig.~\ref{fig:spec}, the $\chi^2$ from the cross-correlation of the
spectra with an elliptical galaxy template spectrum is shown. The
measured values of the redshifts of the two galaxies in the slit are
close, $z_1=0.7579\pm 0.0007$ and $z_2=0.7568\pm0. 0009$, and the
redshift determined from the sum of the spectra is
$z=0.7573\pm0.0006$. Since these galaxies are part of the cluster red
sequence and are located close to its X-ray center, this redshift
measurement can be reliably considered as that of the entire galaxy
cluster. This spectroscopic redhift measurement, obtained using the
observations on the 6-m BTA telescope, is in good agreement with the
photometric estimates obtained above.

\section{Cluster mass estimates}

At the redshift of $z=0.7573$, the X-ray flux of the cluster measured
above corresponds to the X-ray luminosity of
$1.20\pm0.11\cdot 10^{45}$~erg~s$^{-1}$. Using calibration of the
X-ray luminosity-mass relation from \cite{2009ApJ...692.1033V}, one
can obtain an estimate of the cluster mass. Comparison of the masses
of rich clusters at redshifts $z>0.3$ obtained using this relation
from the measurements of X-ray luminosities in SRG/eROSITA survey with
masses measured through the observation of Sunyaev-Zeldovich effect in
the Planck survey, taken from second \emph{Planck} Sunyaev-Zeldovich
sources catalogue \citep[\emph{PSZ2},][]{PSZ2}, shows that the mass
estimates from X-ray data are on average 12.8\% higher. The mass of
\cl2305\ cluster, estimated from its X-ray luminosity and scaled to
take into account the difference with the masses measured in the
Planck survey, turns out to be
$M_{500}=(9.03\pm 2.56)\cdot 10^{14} M_\odot$, where the main
contribution to the error is produced by the scatter of the mass --
X-ray luminosity relation. According to the scaling relations from
\cite{2009ApJ...692.1033V}, a cluster of this mass should have a
temperature of $T=11.6\pm 2.1$~keV. Therefore, this cluster appears to
be among only a few of the most massive clusters at redshifts $z>0.6$
(see the discussion below). Using the measurements of the parameters
of galaxy clusters for a sample of relaxed clusters obtained with
Chandra telescope \citep{2006ApJ...640..691V}, it is possible to
obtain an estimate of the radius $R_{500} = 1430\pm 120$\,pc, which
corresponds to the angular size of about $3\arcmin\!.3$.

The cluster \cl2305\ was not detected in Planck all-sky survey trough
the observation of the Sunyaev-Zeldovich effect
\citep[SZ,][]{1972CoASP...4..173S} above the threshold of the
\emph{PSZ2} catalogue. However, at lower significance level, SZ source
can be detected in Planck survey comptonization parameter maps
processed as it was done in \cite{br17_pazh}. This cluster was found
earlier in Atacama Cosmological Telescope
\citep[ACT][]{2021ApJS..253....3H}, as well as in South Polar
Telescope \citep[SPT][]{2020ApJS..247...25B} SZ surveys. According to
ACT SZ survey galaxy clusters catalog, the mass of this cluster,
scaled to the masses measured in \emph{PSZ2} catalog, appears to be
$M_{500} = (9.18\pm 1.48)\cdot 10^{14} M_\odot$. The mass of the
cluster measured in SPT SZ survey, also scaled to \emph{PSZ2} masses,
turns out to be equal to
$M_{500} = (7.40\pm 0.83)\cdot 10^{14} M_\odot$. In both cases, the
catalogues of these surveys provide only photometric estimates of the
cluster redshift.

\subsection{Joint analysis of data in the X-ray and microwave bands}

The importance of join analysis of X-ray and microwave
observations of the hot gas in galaxy clusters has been discussed long before the secure detections of the SZ effect became available. On the one hand,  the
X-ray flux from the hot gas in a cluster is determined by the gas emission measure, and, therefore, is
proportional to the gas density squared. On the other hand, the decrement in the
brightness of the cosmic microwave background in the direction of the
cluster due to the SZ effect scales linearly with the gas density. The combination of these two quantities opens
up the possibility to determine the distance to the cluster (see
discussion in \citealt{1980ARA&A..18..537S}).

With the advent of sensitive wide-area surveys in X-rays
(SRG/eROSITA) and the microwave band (Planck, SPT, ACT surveys), the evaluation of cluster parameters based solely on the measurements of $F_X$ and $Y$ are of particular interest \citep[see, for example,][]{2015MNRAS.450.1984C}. This approach does not replace photometric or spectroscopic redshift measurements, but allows one to
immediately obtain rough estimates of the mass of a cluster (or cluster candidate) and use this information to optimize the optical followup program, focusing on the most massive clusters. The specific optimization
procedure will be published separately. For cluster \cl2305\, we know $F_X$, $Y$ and $z$. Therefore, it is interesting to verify the consistency of all approaches.

In case when the X-ray flux $F_X$ and redshift $z$ are known, one can use the luminosity--temperature and luminosity--mass relations to determine the parameters of the cluster. If the X-ray data allow us to reliably measure the temperature $T_X$ and the mass of the gas $M_g$, we can also use such values as, for example,
$Y_X\propto T_X\times M_g$ or directly $M_g$ to estimate the total cluster mass \citep[see,
e.g.,][]{2006ApJ...650..128K,2009ApJ...692.1033V}. Similarly, one can use the total comptonization parameter $Y$ in combination with $z$.

Even if the redshift is not known, one can still obtain some useful information on a cluster. First of all, for clusters in the redshift range $z\sim 0.6\,-\,2$, the mass can be roughly estimated (with the accuracy up to a factor $\sim 2$) directly from the value of the X-ray flux $F_X$ \citep[equation 13 from][]{2015MNRAS.450.1984C}:
$M_{500}\approx 1.2\cdot 10^{14} \left ( F_X/10^{-14} \right)^{0.57} \approx 1.3~10^{15}\, M_\odot$. This relation does not directly use the redshift value, except for the assumption that $z\vargtrsim0.6$. The obtained mass estimate is indeed consistent (within the factor of 2) with the estimate based on the luminosity-mass relation for $a$ known $z$ (see above).

Another interesting possibility is to estimate the cluster redshift using the measurements of $F_X$ and $Y$. In the simplest form, one can use the notion that the value of $R_{XSZ}=F_X/Y$ is well correlated with
$z$ for cluster at $z\vargtrsim 0.6$ 
\citep[][]{2012A&A...543A.102P,2015MNRAS.450.1984C}. For this exercise, the combination of the SRG/eROSITA survey data with the catalogues of the \emph{ACT} telescope \citep{2021ApJS..253....3H} appears particularly attractive due to a large overlap of the survey footprints.

The parameters of the relation between the redshift, X-ray and microwave signals depend on the specific 
procedure used to evaluate these signals. In particular, the ACT telescope catalog provides the value $y_{c, fixed}$, which characterises the amplitude of the
\emph{tSZ} signal from the cluster for a fixed geometry of the spatial filter. The choice of the fixed filter actually means that it is possible to directly compare maps obtained by different telescopes after the convolution with the appropriate filters. It turns out that in this case, the relation between $z$ and the ratio of signals in the X-ray and microwave ranges can be written, for example, in the following form:
\begin{eqnarray}
  \frac{F_{X,fixed}}{y_{c,fixed}}\approx 2.9\times10^{-9} z^{-0.85}~{\rm erg\,s^{-1}\,cm^{-2}},
\end{eqnarray}
where ${F}_{X, fixed}$ --- is the flux in 0.4-2~keV energy range for a fixed spatial filter used to search clusters in the X-ray band. The parameters of this \ec{relation} were calibrated for a set of clusters detected in both X-ray and microwave bands, taking into account the specific flux measurement algorithm. Applying this relation to the cluster \cl2305\
($F_{X, fixed}\approx (1.25 \pm0.15)\,10^{-12}\,{\rm erg\, s^{-1}\,   cm^{-2}}$; $y_{c, fixed}\approx 3.14\,10^{-4}$) gives an estimate of the redshift of $z\approx 0.7\pm0.1$. Note that for the selected
fixed spatial filter, the X-ray flux from the cluster \cl2305\ turns out to be about twice as high as the direct flux measurements discussed above. Despite unavoidable (large) uncertainties in measuring $z$ this way, such estimates help immediately classifying this cluster as a potentially distant object.

\begin{figure}[t]
  \centering

  \smfigure{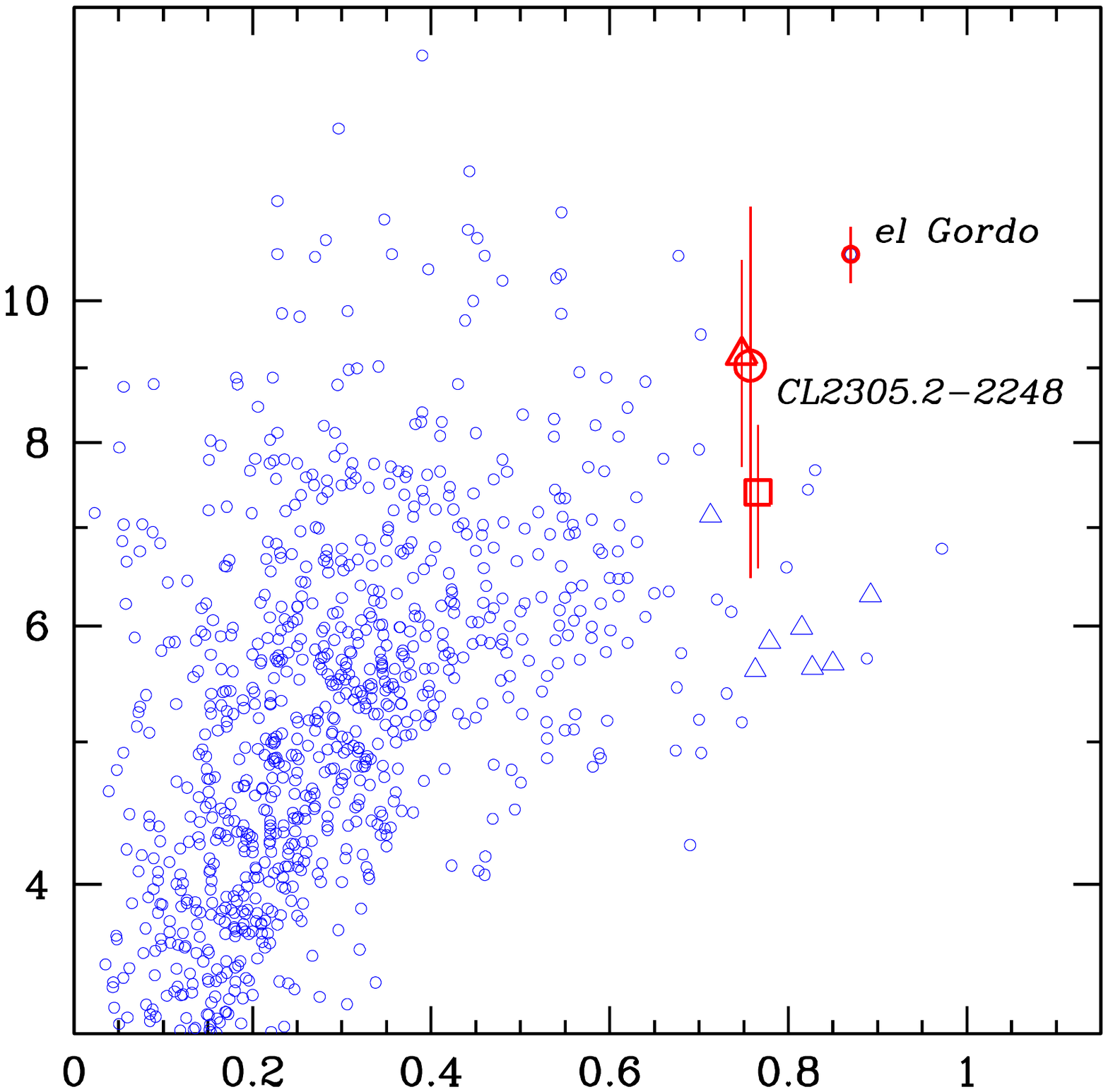}{\emph{z}}{$M_{500}, \times 10^{14} M_\odot$}
  
  \caption{The distribution of galaxy clusters in the redshift -- mass $M_{500}$ plane. The red symbols show the estimates of the mass of
    the cluster \cl2305\ according to SRG/eROSITA (circle), Atacama
    Cosmological Telescope (triangle) and South Polar Telescope
    (square) surveys. For comparison, the red circle also shows the
    mass measurement for the very massive galaxy cluster ``El Gordo''
    \citep{2012ApJ...748....7M}, taken from the 2nd catalog of the
    Planck survey. The blue circles show galaxy clusters from the
    \emph{PSZ2} catalog, the blue triangles --- clusters from the
    \emph{PSZ2} catalog identified in the work of our group
    \citep{br18_pazh}.}
  
  \label{fig:z_m500}
\end{figure}

\section{Discussion}

The very massive galaxy cluster \cl2305\ was detected in the SRG/eROSITA all-sky survey and was identified in optical and IR using the data of \emph{Pan-STARRS} and \emph{WISE} surveys. Observations at the 6-m BTA telescope allow to measure its spectroscopic redshift,
$z=0.7573$, in good agreement with its photometric estimates, including a very accurate one obtained with machine learning methods. Using SRG/eROSITA X-ray data we obtain an estimate of the cluster mass $M_{500}=(9.03\pm2. 56)\cdot 10^{14} M_\odot$. From
mass--temperature scaling relation, the temperature of the cluster can be estimated as $T=11.6\pm 2.1$~keV. The high mass of this galaxy cluster is also confirmed in Atacama Cosmological Telescope SZ survey data, which give a measurement of $M_{500} = (9.18\pm 1.48)\cdot 10^{14} M_\odot$, as well as South Polar Telescope SZ survey data --- $M_{500} = (7.40\pm 0.83)\cdot 10^{14} M_\odot$. All these estimates of the mass of the cluster are scaled to masses measured in second \emph{Planck} Sunyaev-Zeldovich sources catalogue (\emph{PSZ2}).

The central part of the cluster appears rather irregular in the X-ray image. The hot gas in the core could be perturbed by buoyant bubbles inflated by an active galactic nucleus in the central galaxy of the cluster \citep[e.g.,][]{2001ApJ...554..261C} or, alternatively, the core might be affected by a recent merger with a smaller cluster or group. 

The shape of projected distribution of galaxies resembles an hourglass --- in the directions to the east and to the west there are almost no cluster galaxies at distances larger than approximately arcminute from the center of the cluster, while in the direction to the north and south, the surface density of the cluster galaxies remains significant up to distances of the order and greater than the radius $R_{500}\approx 3\arcmin\!.3$. Such a strong central asymmetry of the projected galaxies distribution may indicate that the cluster is a dynamically disturbed system.

Using the data of SRG/eROSITA and ACT surveys in X-ray and microwave bands, and also the cluster redshift measurement, an example of joint analysis of the data from X-ray and microwave observations of the hot
gas in clusters is discussed. We show that cluster \cl2305\ can be identified as a very massive and distant one using the measurements of its X-ray flux and integral comptonization parameter only.

Clusters of galaxies with masses of the order
$M_{500}\approx 10^{15} M_\odot$ are extremely rare objects. This is illustrated in Fig.~\ref{fig:z_m500}, where the distribution of the most massive galaxy clusters in the redshift -- mass $M_{500}$ plane
is shown. One can see that cluster \cl2305\ is found among of only several dozen of the most massive clusters in the observable Universe and among of only a few the most massive clusters of galaxies at $z>0.6$. In this Figure the well-known very massive \emph{El Gordo} cluster \citep{2012ApJ...748....7M} is also shown. Note, that the mass of the cluster \cl2305\ appears to be comparable to the mass of \emph{El Gordo} cluster. Galaxy clusters of that large mass are unique objects in the observable Universe and deserve further detailed
studies.

\acknowledgements

This work is based on observations with the eROSITA telescope on board
the SRG observatory. The SRG observatory was built by Roskosmos in the
interests of the Russian Academy of Sciences represented by its Space
Research Institute (IKI) in the framework of the Russian Federal Space
Program, with the participation of the Deutsches Zentrum f\"ur Luft- und
Raumfahrt (DLR). The SRG/eROSITA X-ray telescope was built by a
consortium of German Institutes led by MPE, and supported by DLR. The
SRG spacecraft was designed, built, launched and is operated by the
Lavochkin Association and its subcontractors. The science data are
downlinked via the Deep Space Network Antennae in Bear Lakes,
Ussurijsk, and Baykonur, funded by Roskosmos. The eROSITA data used in
this work were processed using the eSASS software system developed by
the German eROSITA consortium and proprietary data reduction and
analysis software developed by the Russian eROSITA Consortium.

Observations with the SAO RAS telescopes are supported by the Ministry of Science and Higher Education of the Russian Federation (including agreement No05.619.21.0016, project ID RFMEFI61919X0016). The renovation of telescope equipment is currently provided within the national project ''Nauka''. The authors thank TUBITAK, IKI, KFU and AN RT for support of observations on the Russian-Turkish 1.5-m telescope (RTT-150).

The work was supported by Russian Science Foundation, grant
21-12-00210. AAS was partially supported by the Ministry of Science
and Higher Education of the Russian Federation, project
0033-2019-0005.

\bibliographystyle{astl}
\bibliography{refs_rus}

\end{document}